\documentclass{ws-ijmpa}

\usepackage[super]{cite}
\usepackage{xcolor}
\usepackage[verbose,hypertexnames=false]{hyperref}
\hypersetup{colorlinks=false,allbordercolors=blue,pdfborderstyle={/S/U/W 1}}

\usepackage{overpic,subfigure}
\usepackage{booktabs}

\begin{document}

\markboth{Authors' Names}{Instructions for typing manuscripts (paper's title)}

%
\catchline{}{}{}{}{}
%

\title{Dark photon search status in $\tau-c$ energy region
}

\author{Zhijun Li\footnote{
lizhj37@mail2.sysu.edu.cn}
}
\author{Zhengyun You\footnote{
youzhy5@mail.sysu.edu.cn}
}

\address{School of Physics, Sun Yat-sen University, Guangzhou 510275, China}

\maketitle


\begin{abstract}
The dark photon plays an important role as a portal to dark matter and has been extensively studied on both experimental and theoretical 
frontiers. However, as no signals have been observed, the whereabouts of the dark photon remain a long-standing open question.
With the proposal of a future $\tau-c$ facility, this proceeding reviews and summarizes the current experimental status of the dark photon in the $\tau-c$ energy region, which is expected to provide a reference for future searches. The experimental status indicates that the dark photon in the $\tau-c$ energy region remains highly promising, highlighting the requirement for larger data samples or new methods to further investigate the benchmarks related to the observed cosmic dark matter.
\end{abstract}


\ccode{PACS numbers: 03.65.$-$w, 04.62.+v}

\section{Introduction}
Evidence for the existence of Dark Matter (DM)~\cite{Cirelli:2024ssz} comes from a wide range of astronomical scales, from individual spiral galaxies~\cite{Rubin:1970zza} and clusters of galaxies~\cite{Zwicky:1933gu}, to Cosmic Microwave Background~\cite{Weinberg:2008zzc} and Large Scale Structures at cosmological scales~\cite{Dodelson:2006zt}.
Based on the measurements from cosmology, the most precise nature of DM is the present cosmological DM density, averaged over the whole Universe, with~\cite{pdg2024}
\begin{eqnarray}
\Omega_{\rm{DM}}h^2=0.1200\pm0.0012,
\end{eqnarray}
where $h=0.674\pm0.005$~\cite{Planck:2018vyg} is related to the present Hubble parameter. 

To explain the present DM density, considering DM as a thermal relic by freeze-out is the most plausible mechanism for DM evolution~\cite{Scherrer:1985zt}.
In the freeze-out mechanism, an interaction beyond gravity between DM and SM particles is required. The observed DM density depends on the thermally averaged cross section ($\langle\sigma v\rangle$) of the process $\rm{DM}~\rm{DM} \rightarrow \rm{SM}~\rm{SM}$.
A larger cross-section will create a smaller DM density, and only the appropriate size of cross-section (usually, $\langle\sigma v\rangle=3\times10^{-26}~\rm{cm}^3s^{-1}$) can create the present DM density we measured. The paradigmatic realization of this kind of DM is Weakly Interacting Massive Particles (WIMPs), where the DM mass is TeV-scale and the interaction coupling is similar to the SM weak interaction of $\mathcal{O}(1)$. WIMPs can just result in the observed DM density by lucky coincidence, which has been called the ``WIMP miracle" historically. However, no evidence of WIMP has been observed until now~\cite{XENON:2025vwd}.

Consequently, other possibilities beyond the ``WIMP miracle" have received increasing attention recently. Considering DM with sub-GeV mass, to produce the observed DM density, it requires the presence of light mediators from the dark sector below the weak scale, and the coupling strength between the dark sector and SM particles should be much less than 1.
The annihilation of $\rm{DM}~\rm{DM}\rightarrow\rm{SM}~{SM}$ is mediated by the light mediator, where the light mediator acts as a dark force portal between DM and SM particles. The portal could be vector, scalar, pseudo-scalar, etc. 
The dark photon model is the most natural vector portal model.

Dark photon ($\gamma'$)~\cite{Holdom:1985ag} originates from an extra Abelian gauge group of $U(1)_{\rm{D}}$ (D means Dark), similar to the SM photon ($\gamma$) originating from the SM gauge group of $U(1)_{\rm{Y}}$. The kinetic Lagrangian part of the two Abelian gauge bosons is given by
\begin{eqnarray}
\mathcal{L}=-\frac{1}{4}F_{\mu\nu}F^{\mu\nu}-\frac{1}{4}F'_{\mu\nu}F'^{\mu\nu}-\frac{\epsilon}{2}F'_{\mu\nu}F^{\mu\nu}-\frac{1}{2}m^2_{\gamma'}A'_{\mu}A'^{\mu},
\end{eqnarray}
where $A$ ($A'$) is the spinor of SM (dark) photon, $F^{(')}_{\mu\nu}=\partial_{\mu}A^{(')}_{\nu}-\partial_{\nu}A^{(')}_{\mu}$ is the field strength of $A^{(')}$, $m_{\gamma'}$ is the mass of dark photon, $\frac{\epsilon}{2}F'_{\mu\nu}F^{\mu\nu}$ is the kinetic mixing term, and $\epsilon$ is the mixing strength.
The interaction part of the Lagrangian of the two gauge bosons is given by
\begin{eqnarray}
\mathcal{L'}=eA_{\mu}J^{\mu}+e_{\rm{D}}A'_{\mu}J'^{\mu}+\epsilon eA'_{\mu}J^{\mu},
\label{eq:L'}
\end{eqnarray}
where $e$ is the elementary charge, $e_{\rm{D}}$ is the dark charge, $J^{\mu}=q_f\bar{f}\gamma^{\mu}f$ is the current of ordinary SM matter with the SM fermion charge $q_f$, $J'^{\mu}=\bar{\chi}\gamma^{\mu}\chi$ is the current of dark matter $\chi$.
Lagrangian~(\ref{eq:L'}) describes the interaction between DM and SM particles via the dark photon portal.

\section{Production and decay of the dark photon}
In general, the dark photon can be produced in any process by replacing the SM photon. The production rate is approximately expressed as $\mathcal{R}(\gamma') \sim \epsilon^2 \mathcal{R}(\gamma)$, assuming the phase space differences in the decay are not considered. Here, $\mathcal{R}(\gamma)$ denotes the rate of a photon production process, while $\mathcal{R}(\gamma')$ represents the rate of the corresponding process when replacing $\gamma$ with $\gamma'$ in the final state. 
Traditional searches for the dark photon are typically based on the following production processes:

\begin{itemize}
\item **Bremsstrahlung process**: $e^- Z \to e^- Z \gamma'$ or $p Z \to p Z \gamma'$, which involves using an electron or proton beam to collide with a fixed target. Examples include NA64~\cite{NA64:2018lsq,NA64:2023wbi}, E141~\cite{Riordan:1987aw}, and NA62~\cite{NA62:2023nhs}.
  
\item **Annihilation process**: $e^+ e^- \to \gamma \gamma'$, which can occur in $e^+ e^-$ colliders like BaBar~\cite{BaBar:2014zli,BaBar:2017tiz}, KLOE~\cite{KLOE-2:2018kqf}, and BESIII~\cite{BESIII:2017fwv,BESIII:2022oww}. It can also happen in fixed target experiments, such as NA64~\cite{Andreev:2021fzd}.

\item **Meson decay**: Processes such as $\pi^0/\eta/\eta' \to \gamma \gamma'$, $\phi \to \eta \gamma'$, and $J/\psi \to \gamma' \eta/\eta'$. Relevant experiments include NA48~\cite{NA482:2015wmo}, NA62~\cite{NA62:2019meo}, and BESIII~\cite{BESIII:2018qzg,BESIII:2018aao,BESIII:2024pxo,BESIII:2025otp}.

\item **Drell-Yan process**: $q \bar{q} \to \gamma'$ at the LHC, with examples including LHCb~\cite{LHCb:2019vmc}, CMS~\cite{CMS:2023hwl}, and FASER~\cite{FASER:2023tle}. Additionally, the Bremsstrahlung process and meson decay processes also occur at the LHC.
\end{itemize}

Regarding the decay of the dark photon, we only consider the case in which $2m_e < m_{\gamma'} < 2m_{\tau}$ in this proceeding. In the experimental search, two types of dark photons need to be distinguished:
\begin{itemize}
\item **Visible dark photon**: This scenario requires $m_{\gamma'} \leq 2m_{\chi}$, and the dark photon will decay visibly into SM particles, such as $\gamma' \to e^+ e^-$. The branching fraction (BF) of the decay depends on the mass of the dark photon and the specific decay channel involved.

\item **Invisible dark photon**: This case requires $m_{\gamma'} > 2m_{\chi}$ and $\alpha_{\rm{D}} = \frac{e^2_{\rm{D}}}{4\pi} \gg \epsilon^2 \alpha$ (where $\alpha$ is the fine structure constant). In this scenario, the dark photon will decay invisibly into a DM pair with $\mathcal{B}(\gamma' \to \chi \bar{\chi}) \sim 100\%$.
\end{itemize}

\begin{figure}[htbp] \centering
	\setlength{\abovecaptionskip}{-1pt}
	\setlength{\belowcaptionskip}{10pt}
	\subfigure[] {\includegraphics[width=0.49\textwidth]{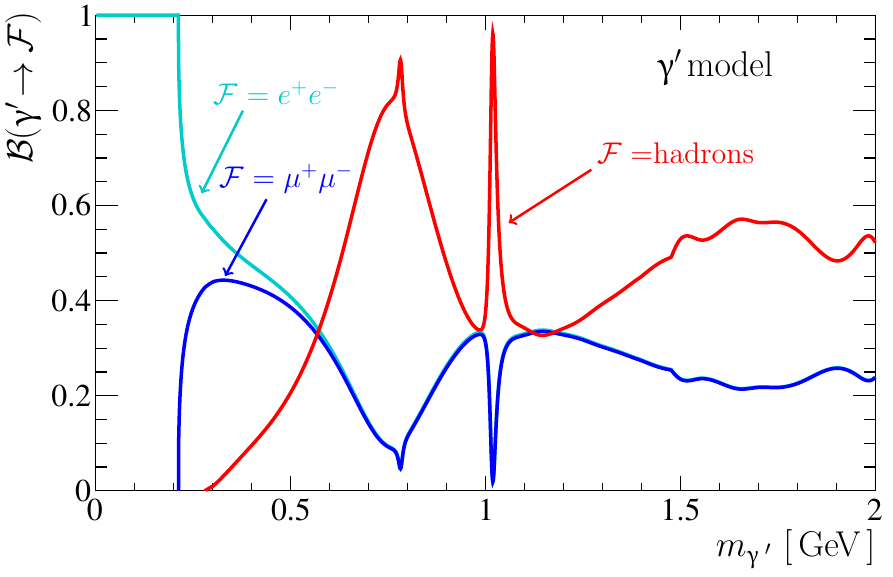}}
    \subfigure[] {\includegraphics[width=0.45\textwidth]{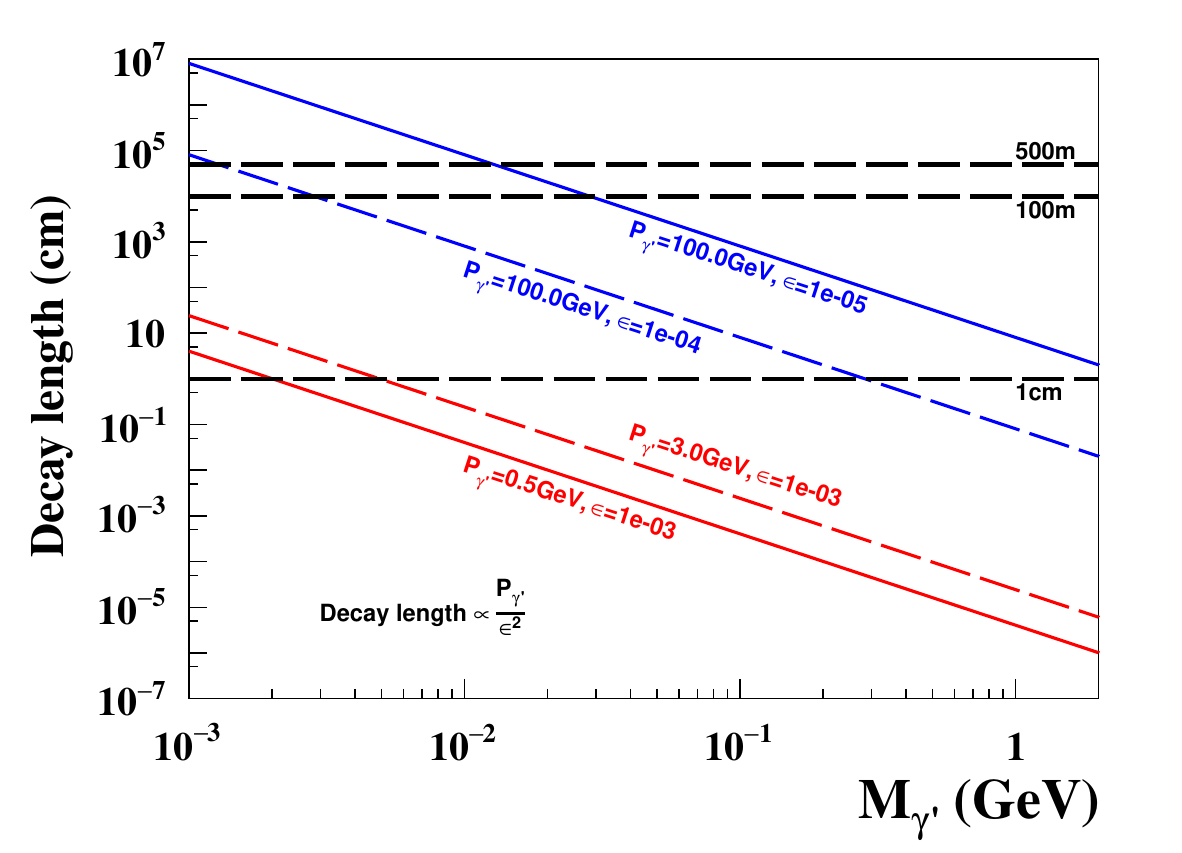}}
	\caption{(a) The BF of $\gamma'\to\mathcal{F}$~\cite{Ilten:2018crw}, where the green line represents $\gamma'\to e^+e^-$, the blue line indicates $\gamma'\to\mu^+\mu^-$, and the red line is $\gamma'\to\rm{hadrons}$. When $m_{\gamma'}\gg m_{\mu}$, $\mathcal{B}(\gamma'\to e^+e^-)\simeq\mathcal{B}(\gamma'\to\mu^+\mu^-)$. (b) Some decay length examples of dark photon with different settings.}
	\label{fig:decay}
\end{figure}

Specifically, the decay width for the processes $\gamma' \to l^+ l^-$ ($l = e, \mu$) and $\gamma' \to \text{hadrons}$ is given by
\begin{eqnarray}
\Gamma(\gamma'\to l^+l^-) = \frac{1}{3}\alpha\epsilon^2 m_{\gamma'} \sqrt{1-\frac{4m^2_{l}}{m^2_{\gamma'}}} \left(1+\frac{2m^2_l}{m^2_{\gamma'}}\right),
\label{eq:W_gp2ll}
\end{eqnarray}
and
\begin{eqnarray}
\Gamma(\gamma'\to\rm{hadrons})=\Gamma(\gamma'\to\mu^+\mu^-)\times R(s=m^2_{\gamma'}),
\label{eq:W_gp2had}
\end{eqnarray}
respectively,
where $\rm{R}(s)=\frac{\sigma(e^+e^-\to\rm{hadrons})}{\sigma(e^+e^-\to\mu^+\mu^-)}$ at the center-mass-energy ($\sqrt{s}$) of $e^+e^-$.
If $m_{\gamma'} \leq 2m_{\chi}$, the total width of $\gamma'$ is given by $\Gamma(\gamma') = \sum \Gamma(\gamma' \to \mathcal{F})$, where $\mathcal{F}$ represents each phase space allowed final state of $l^+l^-$ or $\text{hadrons}$. The BF for the decay $\gamma' \to \mathcal{F}$ is given by $\mathcal{B}(\gamma'\to\mathcal{F})=\frac{\Gamma(\gamma'\to\mathcal{F})}{\Gamma(\gamma')}$, as illustrated in Figure~\ref{fig:decay} (a)~\cite{Ilten:2018crw}.
The decay length of $\gamma'$ is given by $L(\gamma')=\frac{P_{\gamma'}}{m_{\gamma'}}\times\frac{1}{\Gamma(\gamma')}\propto\frac{P_{\gamma'}}{\epsilon^2}$, where $P_{\gamma'}$ is the momentum of $\gamma'$. 
For a dark photon with a small mass and a small mixing strength, it behaves as a long-lived particle. Conversely, for a dark photon with a large mass and a relatively strong mixing strength, it typically decays quickly near the production point (called prompt dark photon). When the dark photon mass and mixing strength fall between these two extremes, it can manifest as a displaced particle. Such characteristics necessitate distinct experimental setups to search for different parameter spaces of the dark photon.
Figure~\ref{fig:decay} (a) shows the decay length of $\gamma'$ with different mass, momentum, and mixing strength.

\vspace{-0.0cm}
\vspace{-0.0cm}

If \( m_{\gamma'} > 2m_{\chi} \), the decay width for the process \( \gamma' \to \chi \bar{\chi} \) is given by
\begin{eqnarray}
\Gamma(\gamma' \to \chi \bar{\chi}) = \frac{1}{3} \alpha_{\rm{D}} m_{\gamma'} \sqrt{1 - \frac{4m^2_{\chi}}{m^2_{\gamma'}}} \left(1 + \frac{2m^2_\chi}{m^2_{\gamma'}}\right).
\end{eqnarray}
The BF for the decay \( \gamma' \to \chi \bar{\chi} \) is given by 
\begin{eqnarray}
\mathcal{B}(\gamma' \to \chi \bar{\chi}) = \frac{\Gamma(\gamma' \to \chi \bar{\chi})}{\Gamma(\gamma' \to \chi \bar{\chi}) + \Gamma(\gamma' \to l^+ l^-) + \Gamma(\gamma' \to \rm{hadrons})}.
\end{eqnarray}
If \( \alpha_{\rm{D}} \gg \alpha \epsilon^2 \), it results that \( \mathcal{B}(\gamma' \to \chi \bar{\chi}) \sim 100\% \) and \( \mathcal{B}(\gamma' \to \rm{SM~Particles}) \sim 0 \). In this case, the dark photon will be ``invisible".
Experimentally, the signatures of visible and invisible dark photons are different, and comparisons or constraints should not be conflated.

\section{Status of the visible dark photon}

On the experimental side, the prompt-visible dark photon is always accompanied by an irreducible background from processes obtained by replacing $\gamma'$ with $\gamma^*$. For instance, in the current search for the process \( e^+e^- \to \gamma \gamma' \) where \( \gamma \to l^+l^- \), there is always the SM background from \( e^+e^- \to \gamma \gamma^* \) followed by \( \gamma^* \to l^+l^- \). In an invariant mass region \( \delta m \) of the final lepton pair, assuming \( \Gamma(\gamma') \ll \delta m \ll m_{\gamma'} \), the cross-section ratio between the dark photon process and the SM process can be expressed as~\cite{Bjorken:2009mm}
\begin{eqnarray}
\frac{\sigma (e^+e^-\to\gamma \gamma'\to \gamma l^+ l^-)}{\sigma(e^+e^-\to \gamma \gamma^*\to \gamma l^+l^-)}=\frac{3\pi\cdot\epsilon^2\cdot m_{\gamma'}\cdot\Gamma(\gamma'\to l^+l^-)}{2\Gamma(\gamma')\cdot\alpha\cdot\delta m},
\end{eqnarray}
and other production processes are also similar.

\begin{figure}[htbp] \centering
	\setlength{\abovecaptionskip}{-1pt}
	\setlength{\belowcaptionskip}{10pt}
	\subfigure[] {\includegraphics[width=0.49\textwidth]{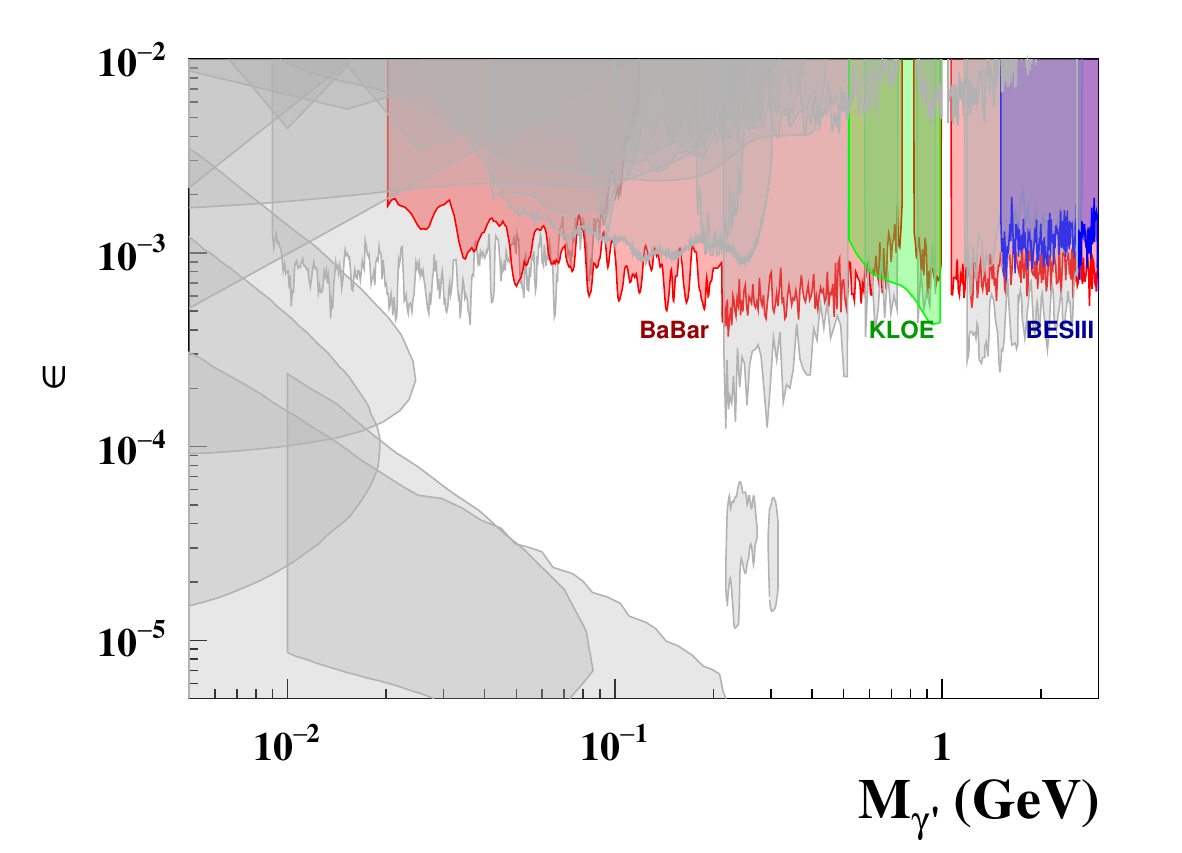}}
    \subfigure[] {\includegraphics[width=0.49\textwidth]{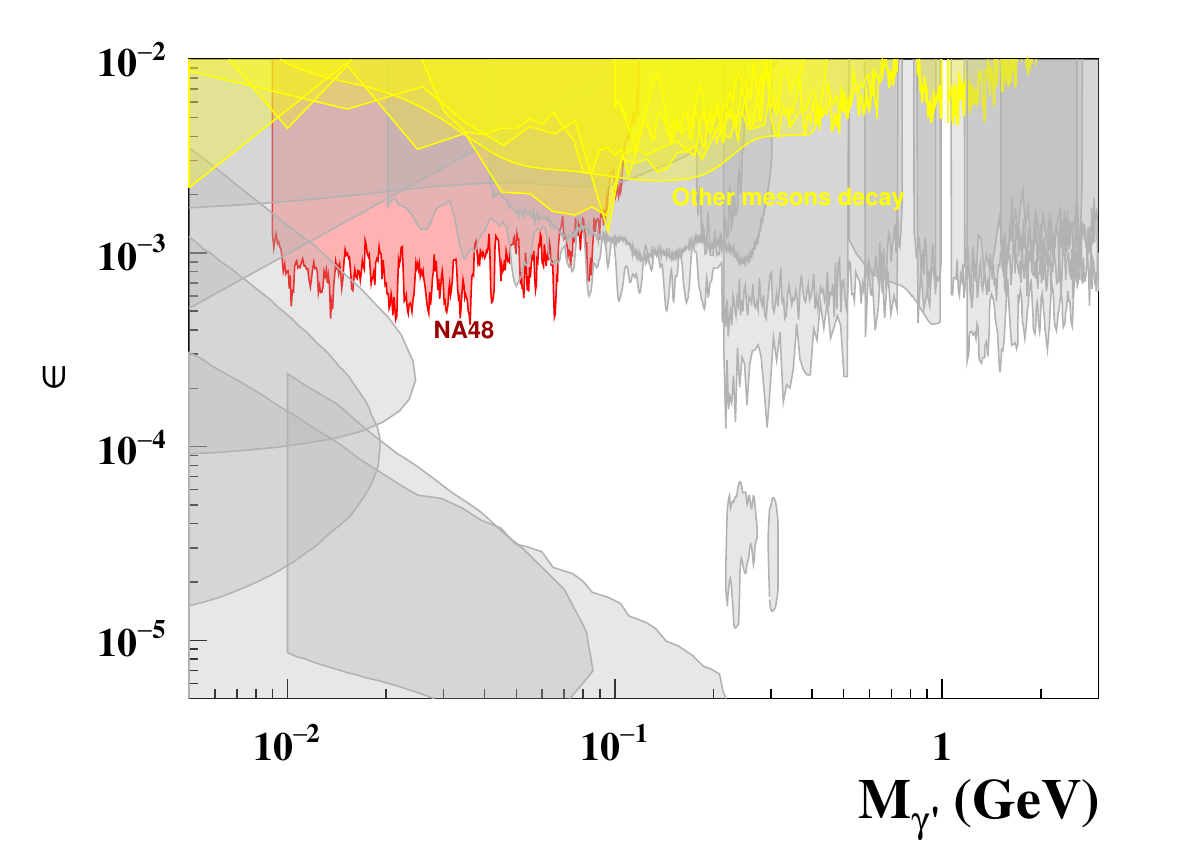}}\\
    \subfigure[] {\includegraphics[width=0.49\textwidth]{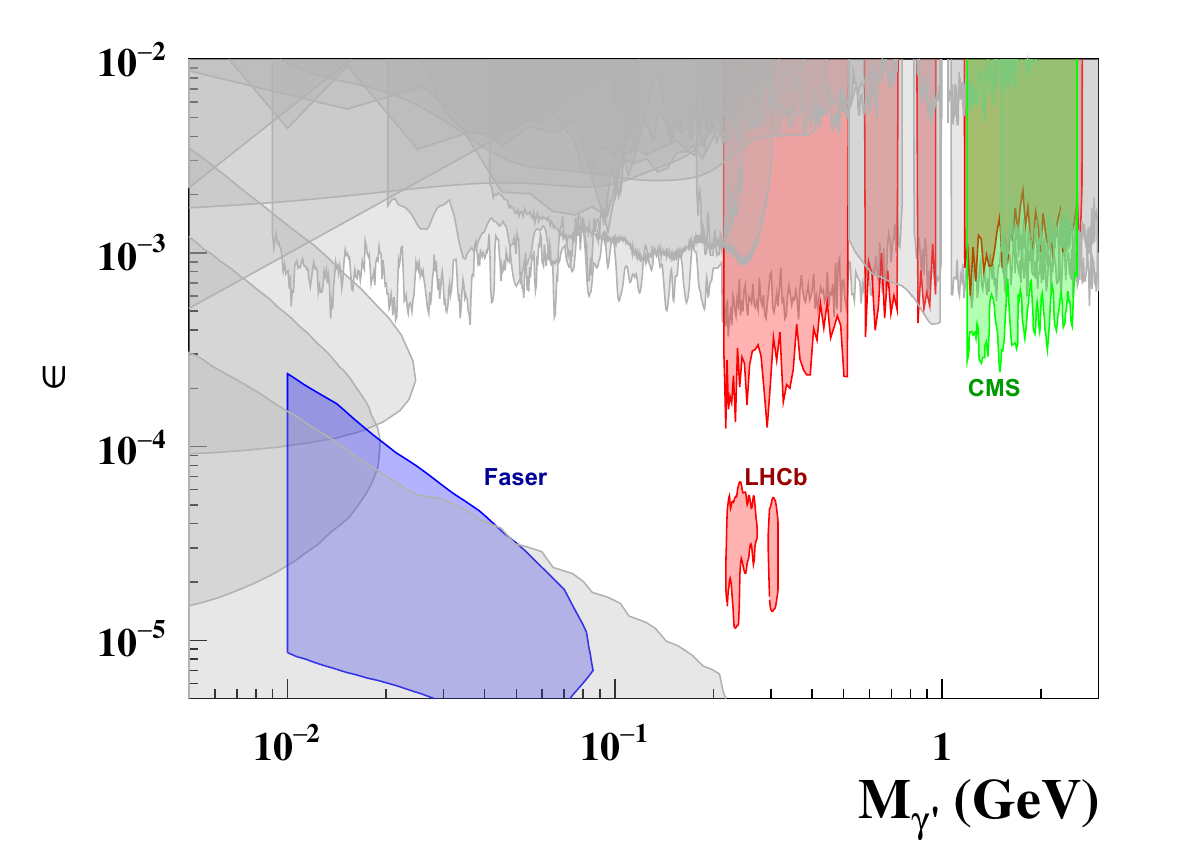}}
    \subfigure[] {\includegraphics[width=0.49\textwidth]{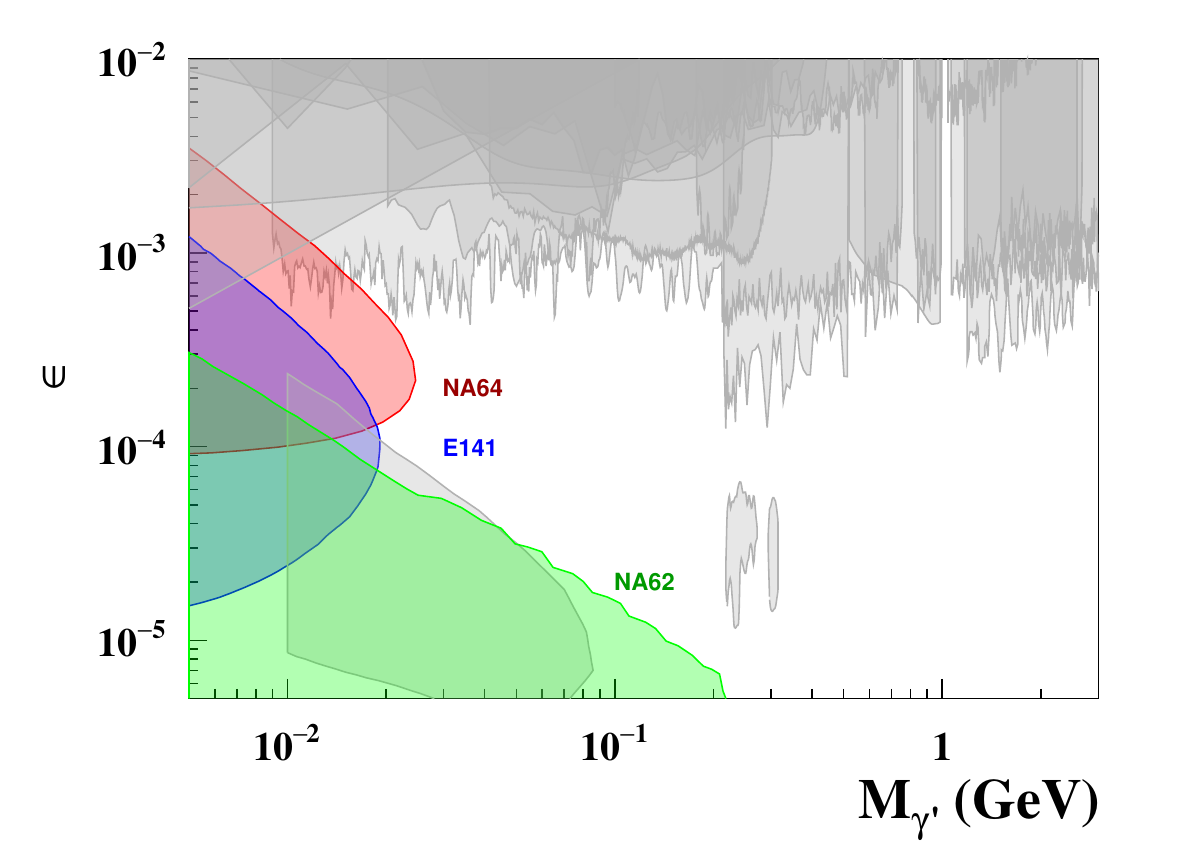}}\\
	\caption{The visible dark photon excluded region from annihilation process (a), meson decays (b), LHC (c), and the fixed target experiment (d).}
	\label{fig:visible}
\end{figure}

The most popular method for searching for the prompt-visible dark photon is the annihilation process $e^+e^-\to\gamma\gamma'$ at the $e^+e^-$ collider, followed by $\gamma'\to X\bar{X}$ ($X=e,\mu,\pi,...$). The cross section of $e^+e^-\to\gamma\gamma'$ is expressed as~\cite{Essig:2009nc}
\begin{eqnarray}
\sigma(e^+e^-\to\gamma\gamma')=\frac{2\pi\epsilon^2\alpha^2}{s}(1-\frac{m^2_{\gamma'}}{s})\left((1+\frac{2m^2_{\gamma'}/s}{(1-m^2_{\gamma'}/s)^2})\Theta-\cos\theta_{\max}+\cos\theta_{\min}\right),
\label{eq:XS_annihilation}
\end{eqnarray}
where $\Theta=\log\left(\frac{(1+\cos\theta_{\max})(1-\cos\theta_{\min})}{(1+\cos\theta_{\min})(1-\cos\theta_{\max})}\right)$ and $[\theta_{\min},\theta_{\max}]$ is the polar angle region between the beam line and the photon momentum.
Some typical searches for $e^+ e^- \to \gamma \gamma'$ are conducted by BaBar~\cite{BaBar:2014zli}, KLOE~\cite{KLOE-2:2018kqf}, and BESIII~\cite{BESIII:2017fwv}. The setups of these three experiments are shown in Table~\ref{tab:annihilation_visible}, and the excluded regions are depicted in Figure~\ref{fig:visible} (a). 
KLOE and BESIII utilize data samples that are two orders of magnitude lower than those of BaBar, yet they achieve competitive constraints on the dark photon, benefiting from the untagged method. From Eq.~\ref{eq:XS_annihilation}, it can be observed that the final gamma photon predominantly distributes around the region of $\cos\theta \sim \pm 1$, which typically corresponds to a blind area in the detector. Although reconstructing the final photon yields improved resolution for $\gamma'$, called the tagged method used in BaBar, this approach results in an extremely low detection efficiency, preventing full utilization of the data's potential. This strongly suggests that an untagged method, similar to that employed by KLOE or BESIII, should be considered for future experimental analyses.

\begin{table}[h]
	\centering
	\caption{The experimental setups of BaBar, KLOE, and BESIII to search for the prompt-visible dark photon in the process $e^+ e^- \to \gamma \gamma'$, where $\mathcal{L}$ represents the integrated luminosity of the data sample used.}
        \begin{tabular}{{c}{c}{c}{c}{c}}
		\midrule \midrule
		Experiment & $\sqrt{s}$ & $\mathcal{L}$ & $\gamma'$ Reconstruction & Method \\
		\midrule
		BaBar~\cite{BaBar:2014zli} & 10.6 GeV & 514 $\rm{fb}^{-1}$ & $e^+e^-,\mu^+\mu^-$ & Tagged \\
        KLOE~\cite{KLOE-2:2018kqf} & 1.019 GeV & 1.93 $\rm{fb}^{-1}$ & $\mu^+\mu^-,\pi^+\pi^-$ & Untagged \\
        BESIII~\cite{BESIII:2017fwv} & 3.773 GeV & 2.93 $\rm{fb}^{-1}$ & $e^+e^-,\mu^+\mu^-$ & Untagged \\
		\midrule \midrule
	\end{tabular}
	\label{tab:annihilation_visible}
\end{table}

\vspace{-0.0cm}
\vspace{-0.0cm}

In addition to the annihilation process, meson decay processes are also essential for studying the prompt-visible dark photon, such as $\pi^0/\eta \to \gamma \gamma'$, $\phi \to \gamma' \eta$, and $J/\psi \to \gamma' \eta/\eta'$, among others. The BF for the decay \( P \to \gamma \gamma' \) (where \( P = \pi^0, \eta, \eta', \ldots \)) is given by
\begin{eqnarray}
\mathcal{B}(P \to \gamma \gamma') = 2\epsilon^2 \left(1 - \frac{m^2_{\gamma'}}{m^2_P}\right)^3 \mathcal{B}(P \to \gamma \gamma),
\end{eqnarray}
and the BF for the decay \( V \to \gamma' P\) (where \( V = \phi, J/\psi, \ldots \)) is given by
\begin{eqnarray}
\mathcal{B}(V \to \gamma'P) = \epsilon^2 |F_{V \to P}(m^2_{\gamma'})|^2 \frac{\lambda^{3/2}(m^2_V, m^2_P, m^2_{\gamma'})}{\lambda^{3/2}(m^2_V, m^2_P, 0)} \mathcal{B}(V \to \gamma P),
\end{eqnarray}
where \( \lambda(a, b, c) = a^2 + b^2 + c^2 - 2ab - 2ac - 2bc \), and \( |F_{V \to P}(m^2_{\gamma'})| \) is the transition form factor from \( V \) to \( P \). This form factor is typically assumed to take a single-pole form given by \( |F(q^2)| = \frac{1}{1 - q^2 / \Lambda^2} \), where \( q^2 \) is the squared four-momentum transfer and \( \Lambda \) is the effective pole mass.
Currently, NA48/2 provides the best constraint on the prompt-visible dark photon in the mass range of $9-70~\rm{MeV}/c^{2}$ by utilizing the meson decay process $\pi^0 \to \gamma \gamma', \gamma' \to e^+ e^-$~\cite{NA482:2015wmo}. NA48/2 uses $K^+$ mesons produced by protons impinging on a target to obtain the $\pi^0$ mesons through the decay $K^+ \to \pi^0$, with a total of $1.69 \times 10^7$ decay candidates of the reconstructed $\pi^0 \to \gamma e^+ e^-$. The excluded region from NA48/2 is shown in Figure~\ref{fig:visible} (b).
There are also results from other meson decays, such as the search for $\phi \to \eta \gamma'$~\cite{KLOE-2:2012lii} at KLOE, and searches for $J/\psi \to \gamma' \eta / \eta'$~\cite{BESIII:2018qzg,BESIII:2018aao}, $\eta / \eta' \to \gamma \gamma'$~\cite{BESIII:2024pxo}, and $\chi_{cJ} \to \gamma' J/\psi$~\cite{BESIII:2025otp} at BESIII. The constraints on the dark photon from these processes are looser than those from $\pi^0 \to \gamma \gamma'$ or $e^+ e^- \to \gamma \gamma'$ due to limited statistics, which are also depicted in Figure~\ref{fig:visible} (b).
However, it is essential to note that in the dark photon model, the vector portal shares a universal coupling strength of $q_f \epsilon$ for different fermions, and the search in meson decays can uniquely probe the coupling with specific quarks in certain non-universal coupling models.

At the LHC, the range of physics processes is abundant and complex, allowing for the production of dark photons through various mechanisms with substantial statistics. These include meson decays such as $\pi^0/\eta \to \gamma \gamma'$, the bremsstrahlung process $pp \to pp \gamma'$, and the Drell-Yan process $q \bar{q} \to \gamma'$, among others. LHCb and CMS have conducted searches for the prompt-visible dark photon by looking for narrow peaks in the dimuon mass spectrum~\cite{LHCb:2019vmc,CMS:2023hwl}. As previously mentioned, these prompt searches must account for the irreducible background from the process $\gamma^* \to \mu^+ \mu^-$.
Unlike the prompt dark photon, displaced and long-lived dark photons necessitate a small mixing strength, which leads to a reduced production rate of the dark photon. However, the large statistics available at the LHC enable the search for displaced or long-lived dark photons to be feasible.
Since the displaced dark photon decay $\gamma' \to \mu^+ \mu^-$ has a different decay vertex compared to the prompt-like process $\gamma^* \to \mu^+ \mu^-$, LHCb can search for the displaced dark photon using the same data sample while experiencing less background. This allows LHCb to explore smaller mixing strengths for the displaced dark photon~\cite{LHCb:2019vmc}.
FASER has also searched for long-lived dark photons produced in proton-proton collisions at the ATLAS interaction point~\cite{FASER:2023tle}. The dark photon travels in the far-forward direction over a distance of 480 m and decays to $e^+ e^-$. Since the SM process $\gamma^* \to e^+ e^-$ is effectively eliminated during its travel, particularly by passing through 100 m of rock, this search represents a background-free analysis. The excluded regions for the visible dark photon from LHC searches are shown in Figure~\ref{fig:visible} (c).

To search for the long-lived dark photon, a key aspect of the experiment is the need for large statistics to ensure a possible signal yield despite the dark photon’s small production rate. Fixed-target experiments are particularly suitable for this task. 
In the experimental setup, a shield is typically placed behind the target to eliminate the SM background from $\gamma^* \to e^+ e^-$. The distance of the shield is denoted as $L_{\rm{sh}}$. The long-lived dark photon can pass through the shield and decay before reaching the detector. A certain decay distance ($L_{\rm{dec}}$) is allocated for the dark photon to decay before the detector. 
Experiments with different beam energy, varying values of $L_{\rm{sh}}$, and distinct $L_{\rm{dec}}$ values are sensitive to dark photons with different mass-mixing parameter combinations ($m_{\gamma'} - \epsilon$). Probing the long-lived dark photon within specific parameter spaces typically requires tailor-made experimental setups. The configurations of several typical fixed-target experiments are shown in Table \ref{tab:FixedTarget_visible}, and their excluded regions are presented in Figure~\ref{fig:visible} (d). These searches always result in an upper bound, as dark photons above this threshold will decay prematurely and be eliminated by the shield.

\begin{table}[h]
	\centering
	\caption{The experimental setups of NA64, E141, and NA62 to search for the long-lived visible dark photon in the processes $e^-Z\to e^-Z\gamma'$ or $pZ\to pZ\gamma'$ with $\gamma'\to e^+e^-$, where $N_{\rm{sig}}$ represents observed signal yield.}
        \begin{tabular}{{c}{c}{c}{c}{c}{c}{c}}
		\midrule \midrule
		Exp. & Beam & Energy of beam & $N_{e^-/p}$ & $L_{\rm{sh}}$ & $L_{\rm{dec}}$ & $N_{\rm{sig}}$ \\
		\midrule
		NA64~\cite{NA64:2018lsq} & $e^-$ & 100 GeV & $5.4\times10^{10}$ & short & 3.5 m & 0 \\
        E141~\cite{Riordan:1987aw} & $e^-$ & 9 GeV & $2\times10^{15}$ & 0.12 m & 35 m & $1126^{+1312}_{-1126}$ \\
        NA62~\cite{NA62:2023nhs} & $p$ & 400 GeV & $2\times10^{17}$ & 100 m & 120 m & 0 \\
		\midrule \midrule
	\end{tabular}
	\label{tab:FixedTarget_visible}
\end{table}

\section{Status of the invisible dark photon}

There are two methods to search for the invisible dark photon in experiments. The first method involves measuring the four-momenta of the final SM particles and calculating the recoiling mass of these particles, known as the missing mass method. In this approach, the dark photon will manifest as a narrow peak in the missing mass distribution.
The second method is to measure the deposited energy of the dark photon in the calorimeter, referred to as the missing energy method. Since the dark photon interacts minimally with SM materials, it does not deposit detectable energy in the calorimeter.

\begin{figure}[htbp] \centering
	\setlength{\abovecaptionskip}{-1pt}
	\setlength{\belowcaptionskip}{10pt}
	\subfigure[] {\includegraphics[width=0.49\textwidth]{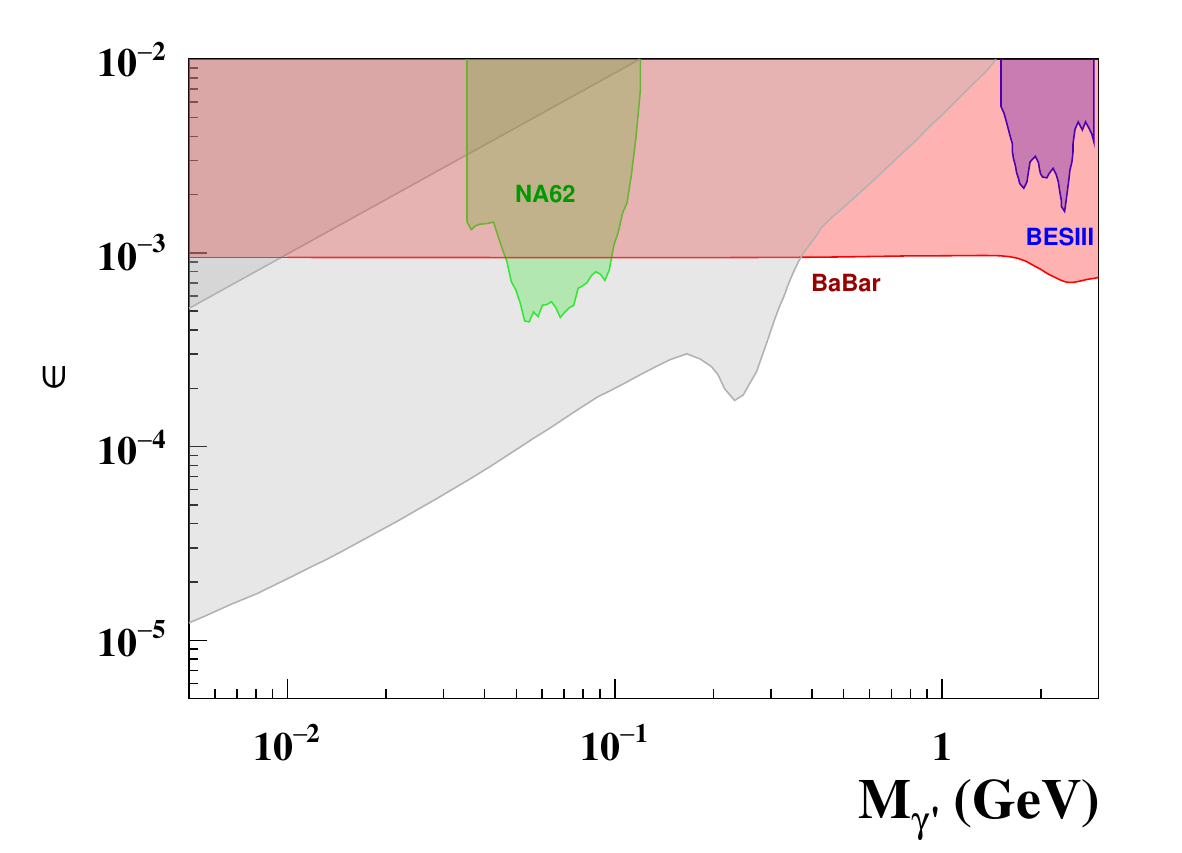}}
    \subfigure[] {\includegraphics[width=0.49\textwidth]{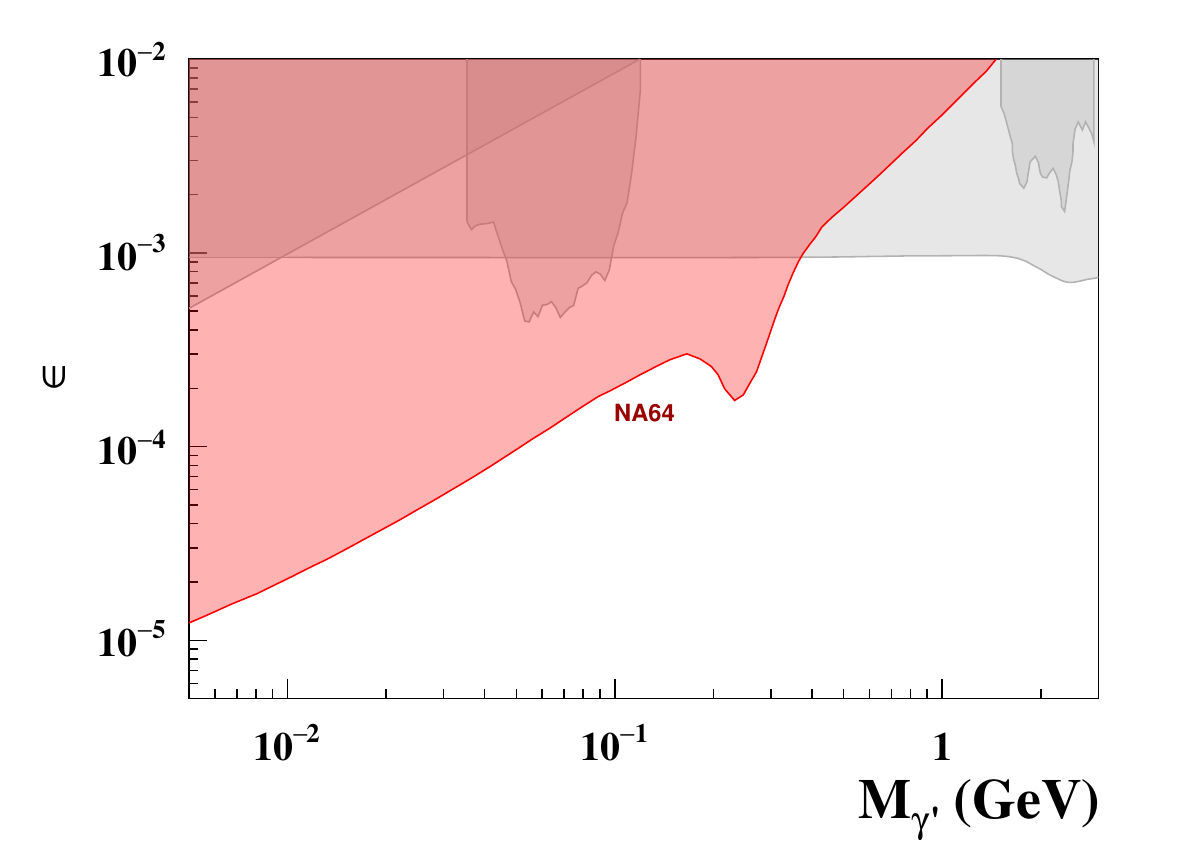}}\\
	\caption{The invisible dark photon excluded region from annihilation process and meson decays by using the missing mass method (a), and the fixed target experiment by using the missing energy method (b).}
	\label{fig:invisible}
\end{figure}

Based on the missing mass method, BaBar, BESIII, and NA62 have conducted several searches. BaBar utilizes $53~\rm{fb}^{-1}$ of $e^+ e^-$ collision data to search for the invisible dark photon produced in the annihilation process $e^+ e^- \to \gamma \gamma'$, providing the leading constraints for $m_{\gamma'} \gtrsim 300~\rm{MeV}$~\cite{BaBar:2017tiz}. BESIII has also performed a similar search using $14.9~\rm{fb}^{-1}$ of $e^+ e^-$ collision data and obtained a competitive constraint~\cite{BESIII:2022oww}.
NA62 analyzes $4 \times 10^8$ tagged $\pi^0$ mesons from $K^+ \to \pi^+ \pi^0$, searching for the invisible dark photon in the decay $\pi^0 \to \gamma \gamma'$. This approach yields better constraints than the annihilation process in some low mass regions~\cite{NA62:2019meo}. 
The excluded regions derived from the missing mass method are shown in Figure~\ref{fig:invisible} (a).

\vspace{-0.0cm}
\vspace{-0.0cm}

Based on the missing energy method, NA64 has provided significantly better constraints on the invisible dark photon in the low mass region~\cite{NA64:2023wbi}, primarily due to the larger statistics available in the fixed target experiment. NA64 utilizes $9.37 \times 10^{11}$ electrons with an energy of 100 GeV to collide with the target, searching for the invisible dark photon in the Bremsstrahlung process $e^- Z \to e^- Z \gamma'$.
In the fixed target experiment, the annihilation process \( e_{\rm{sec}}^+ e_{\rm{ato}}^- \to \gamma' \) warrants special attention, where \( e_{\rm{sec}}^+ \) is the secondary positron resulting from the interaction of the input \( e^- \) and the target, and \( e_{\rm{ato}}^- \) refers to the atomic electrons. Benefiting from the resonance enhancement of \( e_{\rm{sec}}^+ e_{\rm{ato}}^- \to \gamma' \), this process can provide further constraints in the mass region around \( m_{\gamma'} \sim \sqrt{2 E_{e^+} m_{e^-}} \).
The constraints from NA64 are shown in Figure~\ref{fig:invisible} (b).

\section{Benchmark of the thermal relic DM}

\begin{figure}[htbp] \centering
	\setlength{\abovecaptionskip}{-1pt}
	\setlength{\belowcaptionskip}{10pt}
	\subfigure[] {\includegraphics[width=0.49\textwidth]{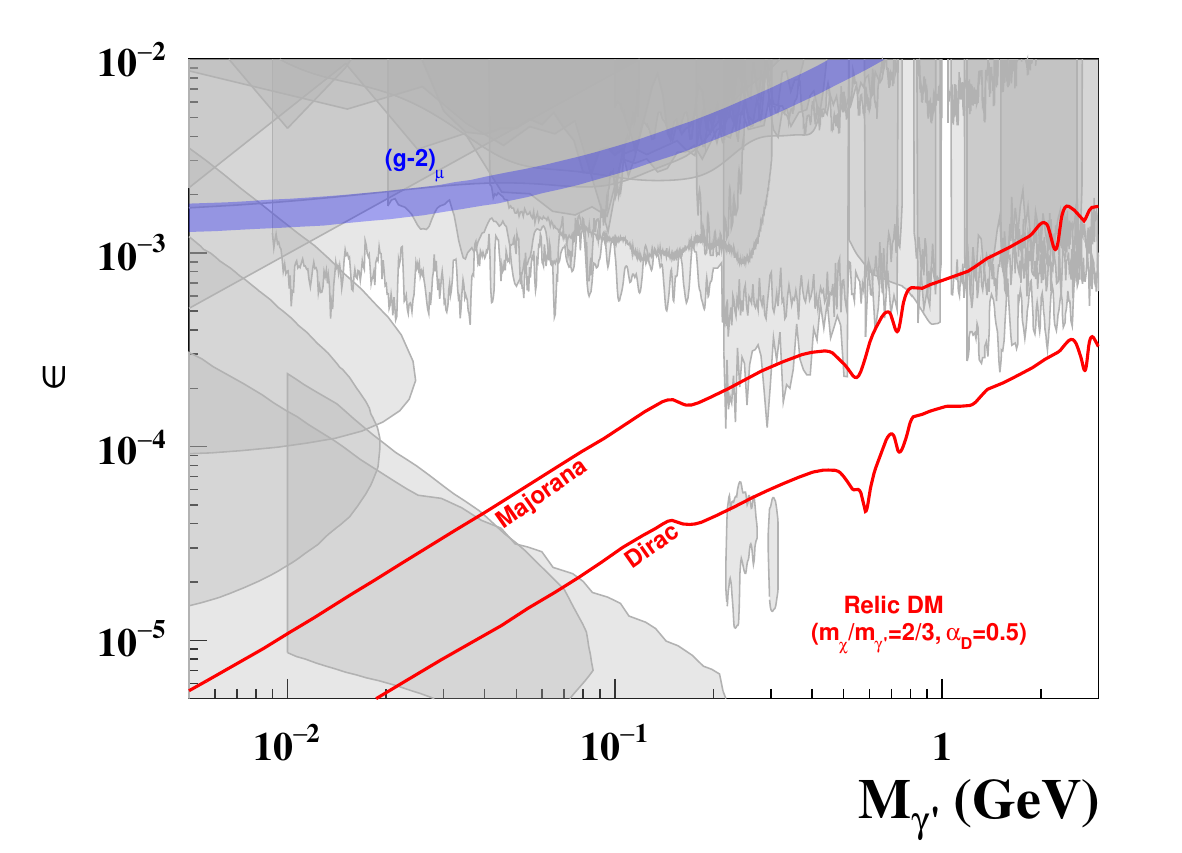}}
    \subfigure[] {\includegraphics[width=0.49\textwidth]{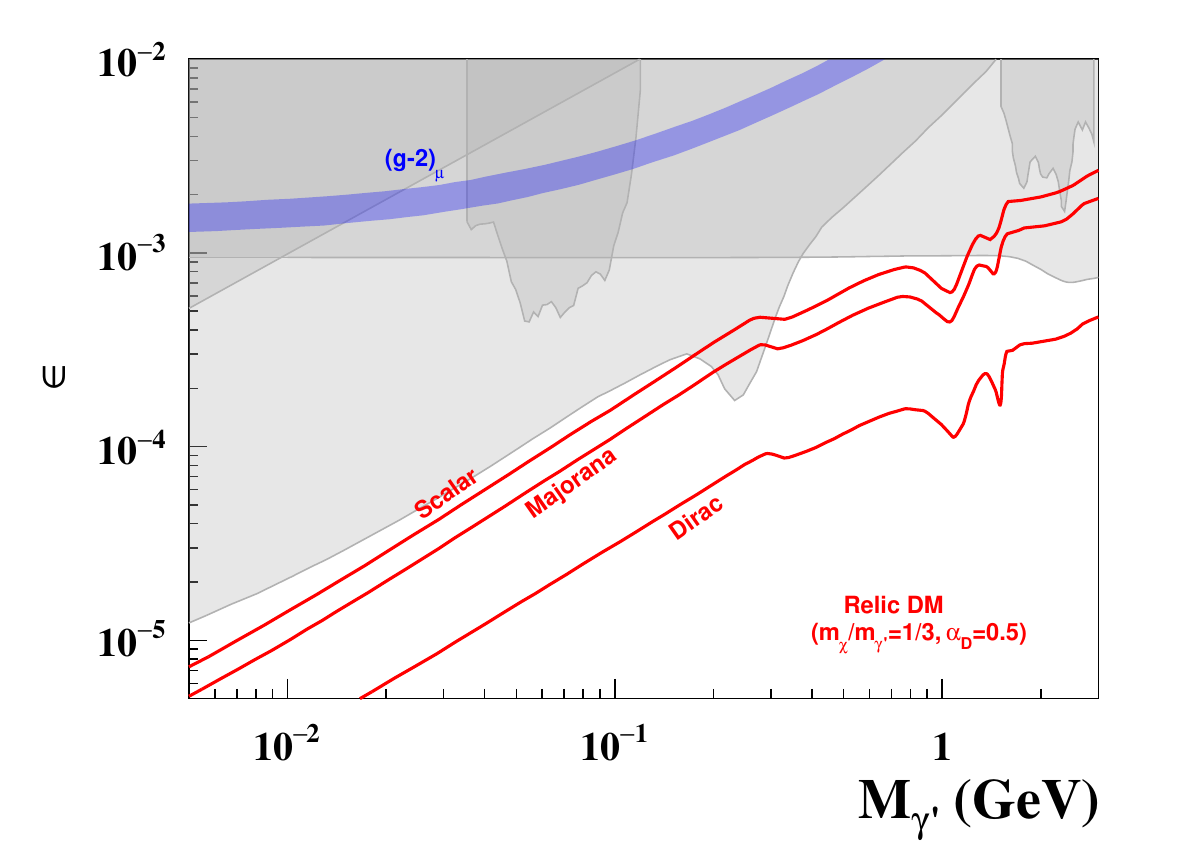}}\\
	\caption{The experimental status on the visible dark photon (a) and invisible dark photon (b). The grey-filled regions represent the experimental excluded region at the 90\% confidence level. The red lines are the benchmark of the observed DM density~\cite{Berlin:2018bsc}, with different DM hypotheses. The blue-filled region is the favored region of the $(g-2)_{\mu}$ anomaly before 2025, which has been excluded and will not be discussed further in this proceeding.}
	\label{fig:DM}
\end{figure}

In the freeze-out mechanism introduced previously, the observed DM in the universe can be explained as thermal relic DM, and the relic density is determined by the thermally averaged cross-section of the process $\chi \bar{\chi} \to \gamma' \to \mathcal{F}$. In the dark photon model, the thermally averaged cross-section is expressed as $\langle \sigma v \rangle \propto \epsilon^2 \alpha_{\rm{D}} \cdot f(m_{\gamma'}, m_{\chi})$, where $f(m_{\gamma'}, m_{\chi})$ is a function of the masses $m_{\gamma'}$ and $m_{\chi}$.
To obtain the observed DM density in the universe, consider the canonical value of $\langle \sigma v \rangle = 3 \times 10^{-26}~\rm{cm}^3 \rm{s}^{-1}$. This allows for the determination of the benchmark for thermal relic DM in the four-dimensional parameter space of $(m_{\gamma'}, \epsilon, m_{\chi}, \alpha_{\rm{D}})$.
Figure~\ref{fig:DM} shows the thermal relic DM benchmarks for the visible dark photon with $\frac{m_{\chi}}{m_{\gamma'}} = \frac{2}{3}, \alpha_{\rm{D}} = 0.5$, and the invisible dark photon with $\frac{m_{\chi}}{m_{\gamma'}} = \frac{1}{3}, \alpha_{\rm{D}} = 0.5$ in the $\epsilon - m_{\gamma'}$ parameter space, along with different DM hypotheses~\cite{Berlin:2018bsc}. The figure also includes a comparison with the currently excluded experimental regions.
One can observe that there are still significant areas that have not been explored by experiments. This situation presents an extremely attractive opportunity and strongly underscores the need for future experiments with larger data samples and new methods to further investigate these benchmarks.

\vspace{-0.0cm}
\vspace{-0.0cm}

\section{Summary}
In this proceeding, we introduce the motivation, search logic, and current experimental status of the dark photon in the $\tau-c$ energy region. If the observed DM in the universe is indeed thermal relic DM produced via the freeze-out mechanism involving the dark photon, we may be on the verge of discovering dark photons or dark matter. The ``freeze-out" mechanism, which is the most natural model for the evolution of DM, combined with the ``dark photon", the simplest extension of the SM, makes this prospect highly promising.
Current experimental data are insufficient to provide a comprehensive exploration of the dark photon, strongly emphasizing the need for future larger data samples, such as a $300 \times$ increase in data at the Super Tau-Charm Facility~\cite{Achasov:2023gey}. 
However, we must also recognize that improvements in statistics alone will not suffice. In most production processes of dark photons, the production rate is directly proportional to $\epsilon^2$. This means that achieving a $10^1 \times$ improvement in $\epsilon$ requires a $10^2 \times$ improvement in the BF or cross-section, and consequently a $10^4 \times$ improvement in statistics under the same analysis strategy. 
Thus, seeking new signatures and developing new analytical methods will also be essential for the future study of dark photons.

\section*{Acknowledgments}

This work is supported in part by National Natural Science Foundation of China (NSFC) under Contracts Nos. 125B2107; National Key R\&D Program of China under Contracts Nos. 2023YFA1606000.




\bibliographystyle{ws-ijmpa}
\bibliography{sample}

\end{document}